# On the security of the Kirchhoff-law–Johnson-noise (KLJN) communicator

Laszlo B. Kish [1] and Claes G. Granqvist [2]

[1] Department of Electrical and Computer Engineering, Texas A&M University, College Station, TX 77843-3128, USA

[2] Department of Engineering Sciences, The Ångström Laboratory, Uppsala University, SE-75121 Uppsala, Sweden

**Abstract** – A simple and general proof is given for the information theoretic (unconditional) security of the Kirchhoff-law–Johnson-noise (KLJN) key exchange system under practical conditions. The unconditional security for ideal circumstances, which is based on the Second Law of Thermodynamics, is found to prevail even under slightly non-ideal conditions. This security level is guaranteed by the continuity of functions describing classical physical linear, as well as stable non-linear, systems. Even without privacy amplification, Eve's probability for successful bit-guessing is found to converge towards 0.5—*i.e.*, the perfect security level—when ideal conditions are approached.

**Introduction.** – In today's normal secure communication, the communicating parties A (Alice) and B (Bob) use software tools to generate and share encryption keys [1]. An eavesdropper (Eve) cannot extract this key because her computational power is limited, but sufficient computing power—for example by using a hypothetical quantum computer or its noise-based logic version—would enable her to fully crack the key. Computing technology progresses at a high pace, which means that today's software-based security is only computationally conditional, and furthermore it does not imply *future-proof security*. Indeed Eve is potentially able to crack a recorded key exchange, and thereby imperil the whole system for data exchange in the near future, even if such a task looks hopeless at the moment.

The inherent deficiencies in the present communication systems have led scientists to explore various physical phenomena for secure key exchange, so that the laws of physics would guarantee security [1]. The aim of these endeavors is to implement a key exchange scheme wherein either the exchange cannot be measured/recorded or, if the information is measured/recorded by Eve, the exchange is nil. This situation is referred to as *perfect unconditional security* or *information theoretic security* [2]. However, perfect security is elusive: one can get ever so close but never reach the goal.

Therefore scientists working with physics-based secure key exchange systems have developed special security devises [3–5], such as *statistical distance* measures between the probability distribution characterizing Eve's extracted key and that of a perfectly secure key of the same length. The *trace distance* in *quantum key distribution* (QKD) is a good example even though its particular use is debated in QKD [3–5]. The numerical value provided by these security measures would be zero for a perfectly secure key. *Practically-perfect* unconditional (or information theoretic) security, which is characteristic for QKD, uses measures that exponentially converge towards perfect security for increasing key length.

The impossibility to have a perfectly secure key in a physical system has led to the common practice to speak of *unconditional security* (or information theoretic security) with an understanding that this is only a practically-perfect unconditional security, *i.e.*, the maximum that can be reached in a physical crypto system. We shall proceed in the same way for the rest of this paper.

It was a commonly accepted assumption for years that only QKD would be able to perform unconditionally secure key exchange, and that this scheme can provide unconditional (practically-perfect) security. However this dogma was refuted by the Kirchhoff-law–Johnson-noise (KLJN) secure key exchange scheme [6] which was introduced in 2005 [7] and subsequently experimentally demonstrated [8]. The KLJN scheme is a statistical/physical competitor to quantum communicators; its security is based on the Second Law of Thermodynamics, which implies that the security of the ideal scheme against passive (non-invasive listening/measuring) attacks is as strong as the impossibility to build a perpetual motion machine of the second kind. The security against active (invasive) attacks is—perhaps surprisingly—provided by the robustness of classical physical quantities, which guarantees that these quantities can be monitored *continuously* without destroying their values (which is totally different for the case of quantum physics).

The KLNJ communicator has been studied during several years with regard to active (invasive) attacks as well as passive attacks associated with non-idealities of the elements embodied in this system [9]. The scheme was consistently found to be unconditionally secure, with the definition above, against each of these attacks.

The present paper presents a general proof for the unconditional security of the KLJN system against any type of non-ideality-based attack, even against attacks that are yet unknown. The proof is based on the continuity of functions describing classical physical linear, and stable non-linear, systems.

**The Kirchhoff-law–Johnson-noise secure key exchange system.** – Figure 1 outlines the KLJN secure key exchange system. For each clock period, *i.e.*, duration of a single bit exchange, Alice and Bob connect their randomly chosen resistor, $R_A$ and $R_B$, respectively, to the line. These resistors are randomly selected from the set $\{R_0, R_1\}$, $R_0 \neq R_1$, where the elements represent the 0 and 1 bit values. The Gaussian voltage noise generators—imitating the Fluctuation-Dissipation Theorem and delivering band-limited white noise with publicly agreed bandwidth—represent enhanced thermal (Johnson) noise at a publicly agreed effective temperature $T_{eff}$ (typically $T_{eff} \geq 10^9 K$ [9]). Their noises are statistically independent from each other and from the noise during the former clock period.

In the case of secure bit exchange (*i.e.*, the 01 or 10 situation), the power density spectrum $S(f)$ and the mean-square amplitude $\langle U_c^2 \rangle$ of the voltage $U_c(t)$ on the wire, and the same quantities of the current $I_c(t)$ in the wire, are given as

$$\langle U_{c,01/10}^2 \rangle = \Delta f \, S_{uc,01/10}(f) = 4kT_{eff} \frac{R_0 R_1}{R_0 + R_1} \Delta f \quad , \tag{1}$$

$$\langle I_{c,01/10}^2 \rangle = \Delta f \, S_{ic,01/10}(t) = \frac{4kT_{eff}}{R_0 + R_1} \Delta f \quad , \tag{2}$$

respectively, where $\Delta f$ is the noise bandwidth; further details of the KLJN system are given elsewhere [6,9]. Eve cannot distinguish between the 01 and 10 situations by measuring mean-square values, because both of them lead to the same results as evident from Eqs. (1) and (2), while the 00 and 11 cases yield distinguishable, and hence insecure, bit arrangements.

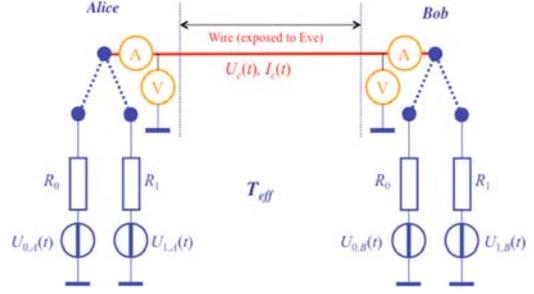

Fig. 1. Schematic of the Kirchhoff-law–Johnson-noise secure key exchange system. The various circuit elements are discussed in the main text. Instantaneous voltage $U_c(t)$ and current $I_c(t)$ amplitudes in the wire are measured by Alice and Bob and compared via a public authenticated data exchange (not shown). $R$, $t$ and $T_{eff}$ denote resistance, time and effective temperature, respectively

The only quantity that potentially could give directional information is the voltage–current cross-correlation, $\langle U_c(t) \, I_c(t) \rangle$ which gives a directional power flow that is zero in thermal equilibrium. Thus the ultimate security of the ideal system against passive attacks is provided by the Second Law of Thermodynamics implying that, for thermal equilibrium, the power $P_{0 \to 1}$, by which resistor $R_0$ is heating resistor $R_1$, is equal to the power $P_{1 \to 0}$ by which $R_1$ is heating $R_0$ [9]. The net power flow between Alice and Bob is zero, as can be quantitatively derived from the Johnson formula, *viz.*,

$$P_{0 \to 1} = \frac{S_{0,uc}(f)\Delta f}{R_1} = 4kT_{eff} \frac{R_0 R_1}{(R_0 + R_1)^2} \Delta f \quad , \tag{3}$$

$$P_{1 \to 0} = \frac{S_{H,uc}(f)\Delta f}{R_0} = 4kT_{eff} \frac{R_0 R_1}{(R_0 + R_1)^2} \Delta f \quad . \tag{4}$$

The equality $P_{H \to L} = P_{L \to H}$, inherent in Eqs. (3) and (4), is in accordance with the Second Law of Thermodynamics, and a violation of this equality would mean going against basic laws of physics and allow Eve to use the voltage–current cross-correlation $\langle U_c(t) \, I_c(t) \rangle$ to extract the bit [9]. This security proof against passive (listening) attacks holds only for Gaussian noise—which has the property that superposition of Gaussian signals remain Gaussian—and its power density spectrum provides the maximum achievable information about the noise. Thus neither higher-order distribution functions nor other tools, such as higher-order statistics, are able to provide additional information.

If Eve is tampering with or changing the system via an active/invasive intervention—such as launching a man-in-the-middle attack [9]—the laws of physics are not enough to guarantee security. Similarly, non-idealities, which represent deviations from the original scheme, pose vulnerabilities. For defending the system against attacks of this kind, the instantaneous voltage and current amplitudes are measured by Alice and Bob, and these quantities are compared via a public au-thenticated data channel (see Fig. 1). Note, due to the classical physical nature of the scheme, Alice and Bob have a full and deterministic model of the system and continuous monitoring of the quantities is allowed. Then, based on their comparison and publicly set preconditions, Alice and Bob decide to keep or discard the bit having compromised security, see below. This authentication uses only $\log_2(M)$ secure bits of the exchanged ones, where $M$ is the number of bits carrying the current and voltage data in the public channel, which shows that authentication is feasible.

**General security proof.** – We now provide a general proof of the unconditional security of the KLJN key exchange. This proof accounts for all non-ideality features, including transients, non-zero propagation time, non-zero cable resistance and capacitance, non-identical temperatures and other inaccuracies. The four steps of the proof are as follows:

(**1**) It has been proven that the ideal KLJN system is perfectly secure [7-9], which means that Eve's probability for making successful guesses of the exchanged key bit is exactly $p = 0.5$, just as for guesses generated by throwing an ideal random coin.

(**2**) Continuous functions describe the variables of passive classical physical systems, such as those in the core KLJN system (and the same holds true even for analog systems in the absence of thresholds and positive feedback). The function $p_\delta(Q)$, describing Eve's probability to make successful guesses for an actual realization of the KLJN system, is also continuous. Here $Q$ is the set of parameters describing the strength of non-ideality, i.e., $Q = \{x_1, x_2, ..., x_k\}$, where the variables $x_i \geq 0$ $(i = 1, 2, ..., k)$ in $Q$ characterize the non-ideal features, and $k$ is a finite number. The parameter $0 \leq \delta \leq \omega$ of the function $p_\delta(Q)$ characterizes the strength of Eve's eavesdropping measurement output; in the case of a wire resistance attack, for example, $\delta$ is the absolute value of the measured difference between the mean-square voltages at Alice's and Bob's side. The upper limit $\omega$ is enforced by Alice and Bob who, due to the classical physical nature of the system, also know Eve's measured $\delta$ and discard bits with $\delta \geq \omega$. They classify the shared bits with $\delta > \omega$ as high-risk and discard them. Alice and Bob could potentially reach perfect security by choosing $\omega = 0$ at arbitrary $Q$, but such a choice would imply that virtually all of the bits would be discarded. Thus a practical system must use a balanced choice of $Q$ and $\omega$ for the best security at given speed, fidelity and cost.

We now show that the perfect security limit can be approached at arbitrary $\omega$ with a proper design of $Q$. For the sake of simplicity, it is easy to define the quantities in $Q$ so that their values are zero in a mathematically ideal case and $p_\delta(Q, \delta)$ is a non-decreasing function of $Q$. Such choices, for example, can be

$x_1$ = cable length, $x_2$ = bandwidth, $x_3$ = resistivity, $x_4$ = 1/cable-diameter, $x_5$ = specific-parasitic-capacitance $x_6$ = propagation-time/transient-protocol-duration, *etc.*

Similarly, it is easy to define $\delta$ so that $p_\delta(Q)$ is a non-decreasing function of $\delta$.

(**3**) In the case of $Q = (0, 0, ..., 0)$, the probability $p_\delta(Q) = 0.5$ (for arbitrary $\delta$) represents perfect security. By using a Taylor expansion at this point, *i.e.*, in the limit $x_i \to 0$ and for all $i$, one obtains

$$0.5 \leq p_\delta(Q) = 0.5 + \sum_{i=1}^{k} \frac{\partial p_\delta [Q = (0,...0)]}{\partial x_i} x_i$$
$$\leq 0.5 + \sum_{i=1}^{k} \frac{\partial p_\omega [Q = (0,...0)]}{\partial x_i} x_i \quad . \quad (5)$$

Thus

$$0.5 \leq p_\delta(Q) = 0.5 + q_\delta(Q) \leq 0.5 + q_\omega(Q) \quad , \quad (6)$$

where $q \to 0$ when $x_i \to 0$ for all $i$, and $q \to 0$ when $\delta \to 0$. In other words, for a given $\omega$ ($\omega > 0$) and by choosing the variables within $Q$ to be non-zero but sufficiently small, it is possible to approach the perfect security limit of $p_a(Q) = 0.5$. A similar type of convergence toward perfect security holds at a given $Q$ ($Q > 0$) for $\omega \to 0$ ($\omega > 0$).

It should be noted that the above security proof does not utilize privacy amplification [10], which can be an additional way to go. Because the bit error probability decays exponentially with increasing duration of the bit sharing period, the remarkably low bit error probability—such as $10^{-12}$ at practical conditions [11] and the resulting extraordinarily high fidelity—allow the use of very high privacy amplification if this is required by economical constrains posed by the values of $Q$ and $\omega$, as further discussed below.

For practical situations, performance aspects (such as speed and wire cost) determine how large magnitudes of $\delta_{max}$ and $Q$ are feasible and how much privacy amplification [10] is required. For example, decreasing $\delta_{max}$ means larger fraction of discarded bits, *i.e.*, slower key exchange speed, while decreasing $Q$ by increasing cable diameter implies a cable cost that grows by the square of this diameter.

(**4**) The above considerations provide a general proof of information theoretic (unconditional) security. Below, we evaluate the security also by a statistical distance measure, namely the *total variation distance* [12] between the probability distribution characterizing Eve's extracted key and that of a perfectly secure key of the same length. We show that this quantity characterizing the security of the key exchange decays exponentially for increasing length of the key.

In order to assess the security of the shared key, one must compare the probability distribution of successfully guessing each possible key sequence of the *N*-bit-long key extracted by Eve, encompassing $2^N$ different sequences, with that of the perfectly secure key having uniform distribution. The total variation distance $\Delta$ [12] of the probability distribution characterizing Eve's extracted key and that of a perfectly secure key of the same length is

$$\Delta(E,I) = \max_{j=1,\ldots,2^N}\left[P(E_j) - P(I_j)\right], \tag{7}$$

where *E* and *I* indicate Eve's extracted key and a perfectly secure key, respectively, and $P(E_j)$ and $P(I_j)$ are the probabilities for correctly guessing the *j*[th] version of Eve's key and of the perfectly secure key, respectively. From Eqs. (6) and (7) we obtain

$$\Delta(E,I) = \max_{j=1,\ldots,2^N}\left[P(E_j) - P(I_j)\right] = \left[0.5 + q_\omega(Q)\right]^N - 0.5^N. \tag{8}$$

For the case of $Nq_\omega(Q) \ll 0.5$, which is a realizable condition according to Eqs. (5) and (6), one then gets

$$\begin{aligned}\Delta &= \left(0.5 + q_\omega(Q)\right)^N - 0.5^N = \\ &= 0.5^N\left[\left(1 + 2q_\omega(Q)\right)^N - 1\right] \cong 2Nq_\omega(Q)0.5^N.\end{aligned} \tag{9}$$

It is thus found that the total variation distance between the probability distribution characterizing Eve's extracted key and that of a perfectly secure key of the same length decays exponentially towards zero for increasing key length provided the condition $q_\omega(Q) \ll 0.5/N$ holds. This condition is given by Eqs. (5) and (6).

It should be noted that this security proof is not valid for non-ideality-based hacking attacks against system components containing active electronics with thresholds, such the current/voltage measurement systems at Alice's and Bob's locations which can be overloaded and saturated by a Makarov-type blinding attack via a large voltage spike [1]. However, a defense against such attacks is easily accomplished by use of proper algorithms that discard the bits when the voltage and current are outside their expected ranges, even if they are identical at Alice's and Bob's sides.

**Conclusion.** – This paper presents a general proof for the unconditional security of the Kirchhoff-law–Johnson-noise communicator against any type of non-ideality-based attack, even against those that are yet unknown. The proof is based on the continuity of functions characterizing classical physical linear, as well as stable non-linear, systems.

**Acknowledgement.** – LK is grateful to Horace Yuen, Vadim Makarov, Renato Renner and Vincent Poor for helpful discussions about relevant security measures.

REFERENCES

[1] GERHARDT I., LIU Q., LAMAS-LINARES A., SKAAR J., KURTSIEFER C. and MAKAROV V., *Nature Commun.*, **2** (2012) 1–6.
[2] LIANG Y, POOR H. V. and SHAMAI S., *Foundations Trends Commun. Inform. Theory* **5** (2008) 355–580.
[3] YUEN H. P., *arXiv*:1109.2675v3 (2012).
[4] HIROTA O., *arXiv*:1208.2106v2 (2012).
[5] RENNER R., *arXiv*:1209.2423v.1 (2012).
[6] ABBOTT D. and SCHMERA G., *Scholarpedia* **8** (2013) 31157. (open acces) http://www.scholarpedia.org/article/Secure_communications_using_the_KLJN_scheme
[7] KISH L. B., *Phys. Lett. A* **352** (2006) 178–182.
[8] MINGESZ R., GINGL Z. and KISH L. B., *Phys. Lett. A* **372** (2008) 978–984.
[9] MINGESZ R., KISH L. B., GINGL Z., GRANQVIST C. G. WEN H., PEPER F., EUBANKS T. and SCHMERA G., *Metrol. Meas. Syst.* **XX** (2013) 3–16. (open access) http://www.degruyter.com/view/j/mms.2013.20.issue-1/mms-2013-0001/mms-2013-0001.xml
[10] HORVATH T., KISH L. B. and SCHEUER J., *EPL* **94** (2011) 28002.
[11] SAEZ Y., KISH L. B., MINGESZ R., GINGL Z. and GRANQVIST C. G., *arXiv*:1309.2179 (2013).
[12] S. ROSENTHAL J. S., *SIAM Rev.* **37** (1995) 387–405.